\documentclass[11pt]{article}
\usepackage{graphicx}
\title{\bf Double inclusive cross-sections for gluon production
in collision of two projectiles on two targets in the BFKL approach}
\author{M.A.Braun\\
Dep. of High Energy physics,
 Saint-Petersburg State University,\\
198504 S.Petersburg, Russia}

\newcommand\lra{\leftrightarrow}
\newcommand\beq{\begin{equation}}
\newcommand\eeq{\end{equation}}

\begin{document}

\maketitle

{\bf Abstract}\\
Double inclusive cross-sections for gluon production in collision of two
nucleons with two nucleons are studied in the BFKL approach.
Various contributions include emission from the pomerons attached
to the participants, from the BFKL interactions between these
pomerons and from the intermediate BKP state. The last contribution
may be observable provided the growth with energy of the pomeron
contribution is tamed in accordance with unitarity. Possibility
of long-range azimuthal correlations due to the BKP state are
discussed.

\section{Introduction}
In our previous papers ~\cite{braun1,braun2} we have derived the forward
scattering amplitude and single inclusive cross-section for gluon production
for two-projectiles-two-target collisions
at high energies in the BFKL approach.
Their immediate applications are to the cross-section for deuteron-deuteron
collisions, although this also concerns a part of
heavy-nucleus-heavy-nucleus collisions due to interaction of two
pairs of nucleons. A remarkable result found in ~\cite{braun1}
is that the cross-sections contain a contribution from the intermediate
state consisting of 4 reggeized gluons in the octet colour state between
the neighbours (BKP state ~\cite{bartels,kwie}). In this paper we study
double inclusive cross-sections for gluon production for the same process.
This process presents a special interest in view of experimental observation
of the long-range azimuthal correlations in the production of a pair of particles.
Note that azimuthal asymmetry in two gluon production has been claimed to be
the source
of the observed correlation in the framework of the JIMWLK
(or Colour Glass Condensate)
approach in collision of two heavy nuclei ~\cite{double}.
We shall study the same problem for
colision of light nuclei, where the results of ~\cite{double} are not valid.
In contrast to the case of heavy nuclei, where only approximate treatment of
heavy nucleus-heavy-nucleus collisions is possible, the light nucleon case allows a
consistent and rigorous study. Thus we are going to see for certain if the initial
process of gluon production really leads to the azimuthal asymmetry.

\begin{figure}
\hspace*{2.5 cm}
\begin{center}
\includegraphics[scale=0.75]{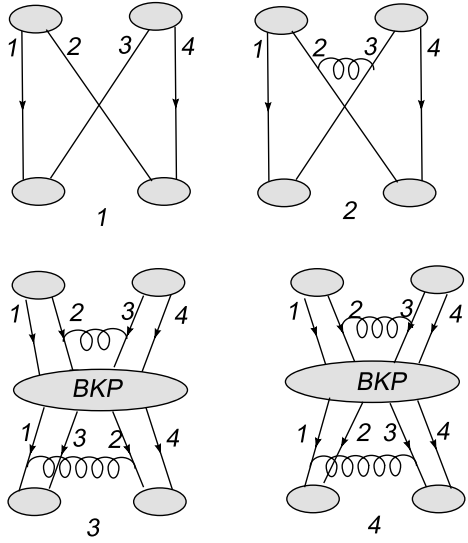}
\end{center}
\caption{ The scattering amplitude for the collision of two projectiles
on two targets}
\label{fig1}
\end{figure}

As derived in ~\cite{braun1} in the lowest order in $\alpha_sN_c$,
assumed small, the scattering amplitude for the collision of two projectiles
on two targets is described by diagrams shown in Fig. \ref{fig1}.

Diagram 1  corresponds to direct sewing of the two pomerons attached
to the projectiles with the two pomerons attached to the targets
with redistribution of colour. Diagram 2  describes the situation when the gluons connecting
the projectiles and targets once interact between themselves.
Diagrams 3 and 4 cover all the rest cases when the exchanged gluons
interact at least twice. The state formed between these interactions is the
BKP state made of 4 reggeized gluons.
The imaginary part of the forward scattering amplitude is the sum
of contributions from Figs. \ref{fig1},1-4:
\beq
D=\sum_{i=1}^4D^{(i)}.
\label{dtot}
\eeq
The forward amplitude $D^{(1)}$ corresponding to Fig. \ref{fig1},1, derived in
~\cite{braun1}, contains an infrared divergent part.
Indeed both of the two pairs of pomerons coupled to the projectiles
and targets contain factor $1/(E-4\omega(q))$ wher $E$ is the two-pomeron
"energy" and $\omega(q)$ is the gluon trajectory, which correspond to free
propagation. The diagram in Fig. \ref{fig1},1  has to contain only one
such factor, which gives an extra factor $E-4\omega(q)$. The term $E$
transforms into $-\partial/\partial Y$ and the divergent part $4\omega(q)$
is cancelled in the sum with terms coming from pomeron interactions in
Fig. \ref{fig1},2.
After this cancellation the infrared finite part of $D^{(1)}$ is given by
\beq
D^{(1)}=-\frac{\partial}{\partial Y}
\int_0^Ydy'\int \frac{d^2q}{(2\pi)^2}P^2(Y-y',q)P^2(y',q).
\label{d1a}
\eeq
Here $P(y,q)$ is the forward pomeron with rapidity $y$ and relative
transversal momentum of reggeized gluons $q$ attached to one of the
participant hadrons. It is the solution of the BFKL  equation for evolution
in rapidity with the initial condition corrsponding to two reggeized gluons
emitted
from the hadron (with their propagators included).
In the following to economize on notations we shall suppress the
transversal variables and integrations over them evident from Fig.
\ref{fig1} and rewrite (\ref{d1a}) as
\beq
D^{(1)}=-\frac{\partial}{\partial Y}
\int_0^Ydy'P^{(12)}(Y-y')P^{(34)}(Y-y')P^{(13)}(y')P^{(24)}(y').
\label{d1}
\eeq
Here we only indicate the numbers of the gluons which combine into pomerons.
In this notation the rest contributions $D^{(i)}$ from Figs.
\ref{fig1},$i$, $i=2,3,4$ were found to be
\beq
D^{(2)}=
\int_0^Ydy'P^{(12)}(Y-y')P^{(34)}(Y-y')(h_{23}+h_{14})P^{(13)}(y')P^{(24)}(y').
\label{d2}
\eeq
\beq
D^{(3)}=
\int_0^Ydy'\int_0^{y'}dy''P^{(12)}(Y-y')P^{(34)}(Y-y')AG(y'-y'')
BP^{(13)}(y'')P^{(24)}(y'').
\label{d3}
\eeq
\beq
D^{(4)}=
\int_0^Ydy'\int_0^{y'}dy''P^{(12)}(Y-y')P^{(34)}(Y-y')A\tilde{G}(y'-y'')
AP^{(12)}(y'')P^{(34)}(y'').
\label{d4}
\eeq
Here
\beq
h_{ik}=-\omega_i-\omega_k-v_{ik}
\eeq
is the BFKL Hamiltonian for gluons $(ik)$; $\omega_i$ is the gluon Regge
trajectory and $v_{ik}$ is the BFKL interaction between gluons $i$ and $k$
in the symmetric form
\beq
v_{ik}(q'_i,q'_k|q_i,q_k)=
4\pi\alpha_sN_c\delta^2(q'_i+q'_k-q_i-q_k)\frac{1}{q'_iq'_kq_iq_k}
\Big(\frac{q_i^2{q'_k}^2+q_k^2{q'_i}^2}{(q_i-q'_i)^2}-(q_i+q_k)^2\Big).
\eeq
Combinations of interactions $A$ and $B$ are given by
\beq
A=v_{23}+v_{14}-v_{13}-v_{24},\ \ B=v_{23}+v_{14}-v_{12}-v_{34},
\label{abdef}
\eeq
both infrared safe. Finally $G$ and $\tilde{G}$ are combinations of the
Green functions for the BKP state for different ordering of the 4 gluons:
\beq
G=\frac{1}{4}[G^{1243}+G^{1342}],\ \
\tilde{G}=\frac{1}{4}[G^{1234}+G^{1432}+G^{1243}+G^{1342}]
\label{gdef}
\eeq
where e.g. $G^{1243}$ satisfies the equation
\beq
\Big(\frac{\partial}{\partial y}-H_{BKP}^{1243}\Big)G^{1243}(y)=\delta(y),
\eeq
with the Hamiltonian
\beq
H_{BKP}^{1243}=-\sum_{i=1}^4\omega_i-v_{12}-v_{24}-v_{43}-v_{31}.
\eeq
In (\ref{d2})-(\ref{d4}) all quantities between the pomerons are to be
understood as operators in the transvserse momentum space. Obviously
transversal integrations include one, two and three intermediate momenta
for (\ref{d1}),(\ref{d2}) and (\ref{d3})+(\ref{d4}) respectively.
Note that in (\ref{d3}) and (\ref{d4}) the interactions in $A$ and $B$
are not to be included into the BKP Green function $G$. The division between
them is related to the change in the colour structure accomplished by these
interactions. Say in the diagram shown in Fig. \ref{fig1},3
the colour structure above
$v_{23}$ and below $v_{12}$ is that of two pomerons and in between is that
of the BKP state, that is of a cylinder with 4 reggeized gluons on
its surface. The interactions $v_{23}$ and $v_{12}$ themselves cannot be
included into either of them.

The observed gluons may be emitted
1) both  from the same pomeron 2) from different pomerons 3) one from a pomeron and the other
from the interactions between the exchanged gluons, both explicitly appearing in the
diagrams of Fig. \ref{fig1} and implicit in the BKP state and finally 4) both from these interactions.
Correspondingly we shall study all contributions successively:
the contributions from the same or different pomerons in Section 2, from a pomeron and
explicitly shown
interactions in Section 3  and both from the interactions and from the
BKP state in Section 4. Section 5 is devoted to some conclusions.

For the double inclusive cross-section in collision of two projectiles with the same momentum
$k$ with two projectiles with the same momentum $l$ the  high-energy part of the forward scattering
amplitude can be presented in the form
\beq
H(Y,y_1,k_1,y_2,k_2)=-i(2\pi)^2\delta(\kappa_+)\delta(q_-)N_c^2(kl)^2 F(Y,y_1,k_1,y_2,k_2),
\label{defh}
\eeq
Here $Y$ is the overall rapidity, $\kappa$ and $q$ are the momenta transferred to the projectile nucleus
and target nucleus respectively with $\kappa_-=\kappa_\perp=q_+=q_\perp=0$.
Factor $N_c^2(kl)^2$ is present in all diagrams, so that it is
convenient to separate it. $F$ gives the
contribution from the diagram for the $S$-matrix
(hence $-i$ in (\ref{defh})). Arguments
$y_1,k_1$ and $y_2,k_2$ are the rapidities and transverse momenta of
the observed gluons. We assume $y_1>>y_2$ motivated by the desire to study long-range correlations in rapidity
(in practice $y_1-y_2\sim 2\div 3)$.
The double inclusive cross-section for two-nucleon-two-nucleon interaction in nucleus-nucleus
scattering  at a given impact parameter $b$ is then
\beq
I_{AB}(Y,b,y_1,k_1,y_2,k_2)\equiv\frac{(2\pi)^2d\sigma_{AB}}{d^2bdy_1d^2k_1dy_2d^2k_2}=\frac{1}{4}
A(A-1)B(B-1)T^{(2)}_{AB}(b) F(Y,y_1,k_1,y_2,k_2),
\label{inclaa}
\eeq
where the transverse density for two pairs of participants is
\beq
T^{(2)}_{AB}(b)=\int d^2b_Ad^2b_BT^2_A(b_A)T^2_B(b_B) \delta^2(b_A-b_B-b)
\eeq
and $T_{A,B}(b)$ are the nuclear profile functions normalized to unity.
For deuteron-deuteron scattering we find instead
\beq
I_{dd}(Y,y_1,k_1,y_2,k_2)=\frac{1}{4}\Big<\frac{1}{2\pi r^2}\Big>_d^2
F(Y,y,k).
\label{incldd}
\eeq
In the following we sometimes suppress the arguments $Y,y_1,k_1,y_2,k_2$ in
our formulas.

After separation of the $\delta$-functions
in Eq. (\ref{defh}) the corresponding diagrams   contain internal
longitudinal integrations over intermediate gluon momenta. For each of the observed
intermediate
gluon these integrations are included in the definition of $I$
leaving only factor $1/4\pi$. Other interactions may refer to either
real gluons in the intermediate state or virtual gluons inside the
production amplitudes. If the gluon is real longitudinal integrations
reduce to the integration over its rapidity with the same
factor $1/4\pi$. If the gluon is virtual integration of
its propagator lifts one of the integrations leaving
the integration over its rapidity with an additional factor
$-i/4\pi$ (see ~\cite{braun1}). Taking into account that
inclusion of the virtual gluon provides an additional factor $(-i)^3$
we find that in the end
virtual integrations give the same result as real ones. This is important for
our calculations: the contribution from the internal gluon line does not depend
on whether it refers to the real gluon (is 'cut') or the virtual one
(is 'uncut'). Apart from integrations over the unobserved or virtual
gluon rapidities, function $F$ is just a contribution from the
diagrams with transverse integrations over the gluon momenta in accordance
with the conservation laws.

\section{Contribution from the pomerons}
\begin{figure}
\hspace*{2.5 cm}
\begin{center}
\includegraphics[scale=0.75]{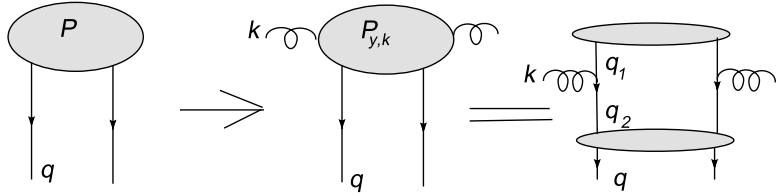}
\end{center}
\caption{'Opening' of the pomeron}
\label{fig2}
\end{figure}

\begin{figure}
\hspace*{2.5 cm}
\begin{center}
\includegraphics[scale=0.75]{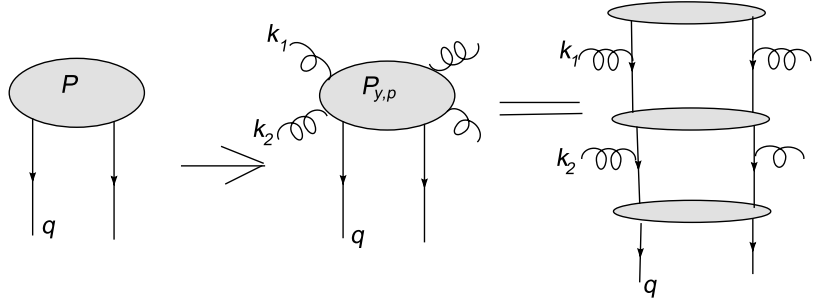}
\end{center}
\caption{Double 'opening' of the pomeron}
\label{fig3}
\end{figure}

For the following note that in our kinematics the pomeron wave function
is real.
The inclusive cross-section for the production of a gluon with rapidity
$y$ and transverse momentum $k$ from the pomeron is well-known. It
corresponds to substitution of the pomeron $P(Y-y',q)$ by the inclusive
cross-section $P_{y,k}(Y,y',q)$ obtained by 'opening' one of the BFKL
interactions inside as shown in Fig. \ref{fig2}.
In our shorthand  notation, suppressing the evident
transversal integrations, we have
\beq
P_{y,k}(Y,y')=
P(Y-y)vG_(y-y'),
\label{ip1}
\eeq
Likewise the double inclusive cross-section corresponds to
substitution of the pomeron $P(Y-y',q)$ by the double inclusive
cross-section $P_{y_1,k_1,y_2,k_2}(Y,y',q)$ illustrated in
Fig. \ref{fig3} and given by
\beq
P_{y_1,k_1,y_2,k_2}(Y,y')=P(Y-y_1)vG(y_1-y_2)vG(y_2-y').
\label{ip2}
\eeq
Note that we assume $y_1>>y_2$, so that diagrams with the  two observed gluons
emitted from the same (vertical) reggeon lines should be summed with those
which contain unobserved gluons emitted at intermediate rapidities,
which convertes these lines into the full BFKL Green function.

The total double inclusive cross-section from the pomerons can be separated into two parts:
with both gluons emitted from the same pomeron and with gluons emitted from different pomerons
from among the four which are present in the diagrams of Fig. \ref{fig1}.

In the first case the double inclusive cross-section will be given by the same formulas as for
the single inclusive cross-section in ~\cite{braun2} in which one only has to substitute
\beq
P_{y,k}(Y,q)\to P_{y_1,k_1,y_2,k_2}(Y,q).
\eeq
So one finds the contribution to $F$
\beq
F^{(1)}=2D(one \ of\ P's\to P_{y_1,k_1,y_2,k_2})
\label{f1}
\eeq
where $D$ is given by (\ref{d1})-(\ref{d4}) and the extra factor two is due
to two possibilities of choosing the production amplitude and its complex
conjugate.

\begin{figure}
\hspace*{2.5 cm}
\begin{center}
\includegraphics[scale=0.75]{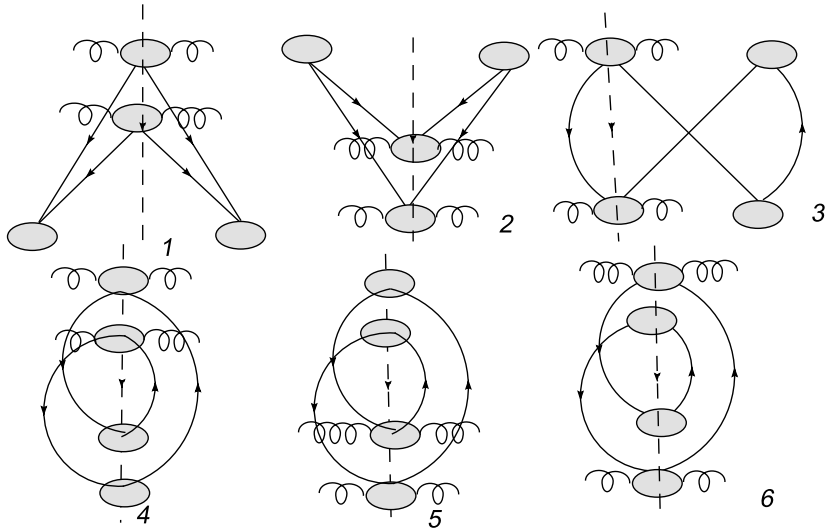}
\end{center}
\caption{Cuts for emission of two gluons from different pomerons}
\label{fig4}
\end{figure}

So we study in some detail only the second case when the two
observed gluons belong to different pomerons.
In this case the double inclusive cross-section
 will be obtained if we cut the diagrams
in all possible ways and substitute two of the cut pomerons according to
Eq. (\ref{ip1}). A particular cut may also pass or not pass through
explicit interactions in Fig. \ref{fig1}, 2-4 or inside the BKP state.
The resulting cross-section from the pomerons will not depend on whether
these extra interactions are cut or not, since the cut BFKL interaction is
equal to uncut one. So one can study different contributions from the
pomerons forgetting about these extra interactions.
The relevant  cuts in this case
 are
shown in Fig. \ref{fig4}.

Diagrams 1 and 2 correspond to diffractive
configurations respective to the target (DT) or projectile (DP). Diagram 3
illustrates the single cut configuration (S) in which one projectile
and one target are cut. Diagrams 4-6  show the double cut configuration
(DC) in which both projectiles and both targets are cut.
Note that if the two gluons are both emitted from the pomerons attached to
the projectile (Fig. \ref{fig4},1) or  both emitted from the pomerons attached to the
target  (Fig. \ref{fig4},2) they can bear any rapidity $y_1$ or $y_2$.
If one of the gluons is emitted from the pomeron attached to the projectile and the
other from the pomeron attached to the target then the first should have rapidity $y_1$
and the second rapidity $y_2$. So effectively the contributions from
Figs. \ref{fig4},1,2 4 and 5 are to be multiplied by two.

We start with the diffractive contributions.
The one respective to the target (Fig. \ref{fig4},1) gives the contribution
to the high-energy part $F$ from Fig. \ref{fig1},1
\beq
F_{DT}^{2,1}=4\frac{\partial}{\partial Y}
\int_0^Ydy'P^{(12)}_{y_1,k_1}(Y,y')P^{(34)}_{y_2,k_2}(Y,y')P^{(13)}(y')P^{(14)}.
\label{fpdt}
\eeq
Coefficient 4 takes into account two projectiles and two targets.
The diffractive contribution respective to the projectile
(Fig. \ref{fig4},2) gives
\beq
F_{DP}^{2,1}=4\frac{\partial}{\partial Y}
\int_0^Ydy'P^{(12)}(Y-y')P^{(34)}(Y-y')P^{(13)}_{y_1,k_1}(y',0)
P^{(24)}_{y_2,k_2}(y',0).
\label{fpdp}
\eeq

The single cut contributions (Figs. \ref{fig4},3)
enter with the minus sign. In fact they contain
one exchanged reggeon on the left and three on the right giving
$i(-i)^3=-1$.
Their contribution is
\beq
F_{S}^{2,1}=-4\frac{\partial}{\partial Y}
\int_0^Ydy'
P^{(12)}_{y_1,k_1}(Y,y')P^{(34)}(Y-y')P^{(13)}_{y_2,k_2}(y',0)P^{(24)}(y').
\label{fps}
\eeq
Coefficient 4 again takes into account interchanges
of the two projectiles and two targets.

Finally  the DC contribution (Figs. \ref{fig4},4, 5 and 6) gives
\[
F_{DC}^{2,1}=4\frac{\partial}{\partial Y}
\int_0^Ydy'\Big(
P^{(12)}_{y_1,k_1}(Y,y')P^{(34)}_{y_2,k_2}(Y,y')P^{(13)}(y')P^{(24)}(y')\]\beq+
P^{(12)}(Y-y')P^{(34)}(Y-y')P^{(13)}_{y_1,k_1}(y',0)P^{(24)}_{y_2,k_2}(y,0')+
2P^{(12)}_{y_1,k_1}(Y,y')P^{(34)}(y-y')P^{(13)}_{y_2,k_2}(y',0)P^{(24)}(y')\Big).
\label{fpdc}
\eeq
In the first two terms the numerical factor is 4 corresponding to different choice of
$(y_1,k_1)$ and $(y_2,k_2)$ in upper or lower pomerons and two  and  two different diagrams
for the DC configuration. In the last term this factor is twice larger since
$(y_1,k_1)$ and $(y_2,k_2)$ can be distributed in four manners between upper and lower
pomerons.

In the sum  the S contribution cancels half of the last term in the
DC contributions  and we
get the final $F$ from the gluons inside the pomerons as
\[
F^{2,1}=4\frac{\partial}{\partial Y}
\int_0^Ydy'\Big(
2P^{(12)}_{y_1,k_1}(Y,y')P^{(34)}_{y_2,k_2}(Y,y')P^{(13)}(y')P^{(24)}(y')\]\beq+
2P^{(12)}(Y-y')P^{(34)}(Y-y')P^{(13)}_{y_1,k_1}(y',0)P^{(24)}_{y_2,k_2}(y,0')
+P^{(12)}_{y_1,k_1}(Y,y')P^{(34)}(y-y')P^{(13)}_{y_2,k_2}(y',0)
P^{(24)}(y',q)\Big).
\label{fp}
\eeq

As mentioned, inclusion of other interactions in between does not change
the form of the result, which is obtained from the  formulas
for the forward amplitude making the substitutions (\ref{ip1}) in the
same manner as above. Referring the reader to ~\cite{braun2} for the details we
only present the final resuls here. For the diagram in Fig. \ref{fig1},2
we find the corresponding $F$ as
\[
F^{2,2}=4\int_0^Ydy'
\Big(
2P^{(12)}_{y_1,k_1}(Y,y')P^{(34)}_{y_2,k_2}(Y,y',q')
HP^{(13)}(y')P^{(24)}(y')\]\[+
2P^{(12)}(Y-y')P^{(34)}(Y-y')HP^{(13)}_{y_1,k_1}(y',0)
P^{(24)}_{y_2,k_2}(y,0')\]\[+
P^{(12)}_{y_1,k_1}(Y,y')P^{(34)}(y-y')HP^{(13)}_{y_2,k_2}(y',0,q')
P^{(24)}(y')\]\beq+
P^{(12)}_{y_1,k_1}(Y,y')P^{(34)}(y-y')HP^{(13)}(y')P^{(24)}_{y_2,k_2}(y',0)
\Big).
\label{fp23}
\eeq

For the diagram in Fig. \ref{fig1},3 with the BKP state
we have the amplitude $F$
\[
F^{2,3}=4\int_0^Y dy'\int_0^{y'}dy''
\Big(2P^{(12)}_{y_1,k_1}(Y,y')P^{(34)}_{y_2,k_2}(Y,y',q_4)M(y'-y'')
P^{(13)}(y")P^{(24)}(y")\]\[+
2P^{(12)}(Y-y')P^{(34)}(Y-y')M(y'-y'') P^{(13)}_{y_1,k_1}(y",0)
P^{(24)}_{y_2,k_2}(y")\]\[+
P^{(12)}_{y_1,k_1}(Y,y')P^{(34)}(Y-y')M(y'-y'')P^{(13)}_{y_2,k_2}(y",0)
P^{(24)}(y")\]\beq+
P^{(12)}_{y_1,k_1}(Y,y')P^{(34)}(Y-y')M(y'-y'')P^{(13)}(y")
P^{(24)}_{y_2,k_2}(y",0)
\Big).
\label{f4p}
\eeq
Here $M=AG(y'-y'')B$ where $A,B$ and $G$ are defined in (\ref{abdef}) and
(\ref{gdef})

The contribution $F^{2,4}$ from the diagram in Fig. \ref{fig1},4 will be given by
a similar expression with the lower pomerons $P^{(13)}P^{(24)}
\to P^{(12)}P^{(34)}$
and $M\to\tilde{M}=A\tilde{G}A$.
The total contribution of emission from two different pomerons is
\beq
F^{(2)}=\sum_{i=1}^4F^{2,i}.
\label{f2}
\eeq

\section {One gluon emitted from the pomeron, the other from interactions}
\subsection{Single interaction between the
pomerons}
\begin{figure}
\hspace*{2.5 cm}
\begin{center}
\includegraphics[scale=0.75]{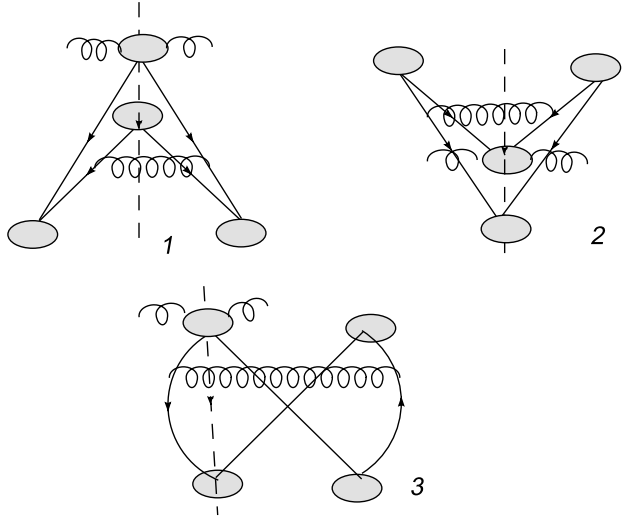}
\end{center}
\caption{Some typical diagrams for emission of one gluon from the pomeron
and another from interaction, with diffractive cuts (1 and 2) and a single
cut (3). The rapidity and momentum of the interaction are assumed to
be fixed either to ($y_1,k_1$) or ($y_2,k_2$)}
\label{fig5}
\end{figure}

Here we study the contribution to the double inclusive cross-section in which
one gluon comes from the 4 pomerons attached to the projectiles or targets and the other
from opening the interaction in the diagram of Fig. \ref{fig1},
2.
Again we shall have different contributions depending on the cuts in
the overall amplitude. Some typical of them are  shown in Fig.
\ref{fig5},1-3 (the full list of diagrams may be inferred from those
in ~\cite{braun2})

Diagrams 1 and 2 describe the two D configurations, DT and DP,
diagram 3 describes the S configuration.
Note that the DC contribution
does not contain any observed gluon in the
intermediate state and so gives no contribution. All contributions
should be taken
with coefficient 4 due to interchanges of projectiles and targets.
which become non-equivalent due to attachment of cut or non-cut pomerons,
Also in the diagrams Fig. \ref{fig5},1 and 3
the gluon emitted from
the pomeron is to have the larger  rapidity than the one emitted from the
interaction and in the diagram Fig. \ref{fig5},2 and the one similar
to Fig. \ref{fig5},3 with emission from the lower pomeron the gluon emitted
from the pomeron is to have the lower rapidity.
Taking into account  contributions from all diagrams we find that the
S contributions cancel 1/2 of the diffractive contributions.
(The extra factor in the S contribution appears because of the fact that
of two possible interactions $v_{23}$ and $v_{13}$ only half of them
contains a real gluon in the intermediate state).
So
for the case when the gluon of higher rapidity is emitted from the pomeron
the result is
\beq
F^{3,h}=4\int_0^Ydy'
P^{(12)}_{y_1,k_1}(Y,y')
P^{(34)}(Y-y')v_{23}P^{(13)}(y')P^{(24)}(y')
\eeq
and for the case when the gluon of higher rapidity is emitted from the
interaction the result is
\beq
F^{3,l}=4\int_0^Ydy'
P^{(12)}(Y-y')
P^{(34)}(Y-y')v_{23}P^{(13)}_{y_2,k_2}(y',0)P^{(24)}(y').
\eeq
The total is
\beq
F^{(3)}=F^{3,h}+F^{3,l}.
\label{f3}
\eeq

\subsection{Two interactions between the
pomerons with
redistribution of colour}
\begin{figure}
\hspace*{2.5 cm}
\begin{center}
\includegraphics[scale=0.75]{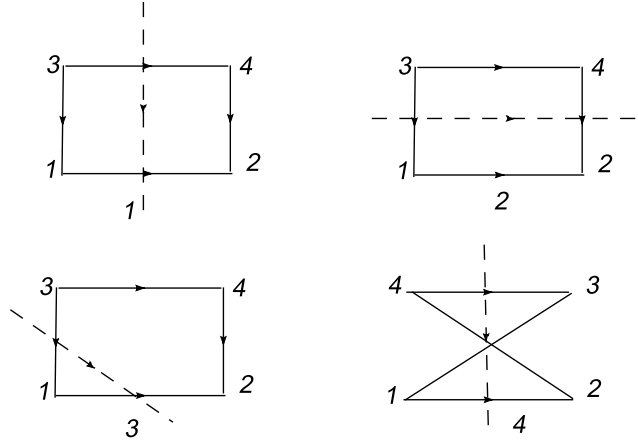}
\end{center}
\caption{Cut interactions between different exchanged gluons for different
configurations}
\label{fig6}
\end{figure}

In this subsection we study contributions to the double inclusive cross-section which
comes from opening one of the pomerons and
one of the interactions explicitly shown in Fig. \ref{fig1},3.
The relevant cuts
are to pass through the interaction which contains the observed gluon.
They have also to pass through the diagram as a whole and thus have to
pass through some pomerons attached to the projectile and target and also
through the BKP state. As a diagram the latter is a cylinder with 4 reggeized gluons arranged
on its surface in order 1243 or 1342  with interaction between neighbours. The cut of the amplitude
generates a cutting plane between reggeized gluons in the BKP state.
As discussed in ~\cite{braun2}, the position of this plane
is totally
determined by the cut passing through the pomerons
and for different configurations is shown in Fig. \ref{fig6} by its
trace on the plane orthogonal to the cylinder.

The forward scattering amplitude is given by (\ref{d3}).
Since cut and uncut interactions give the same contribution, this also
means that in all cases the BKP state will appear as a whole between
the interactions which generate the observed gluon.
So to simplify notations in some of the following formulas we put
$G(y)\to 1$ and  introduce this  Green
function between the interactions only in the end.

\begin{figure}
\hspace*{2.5 cm}
\begin{center}
\includegraphics[scale=0.75]{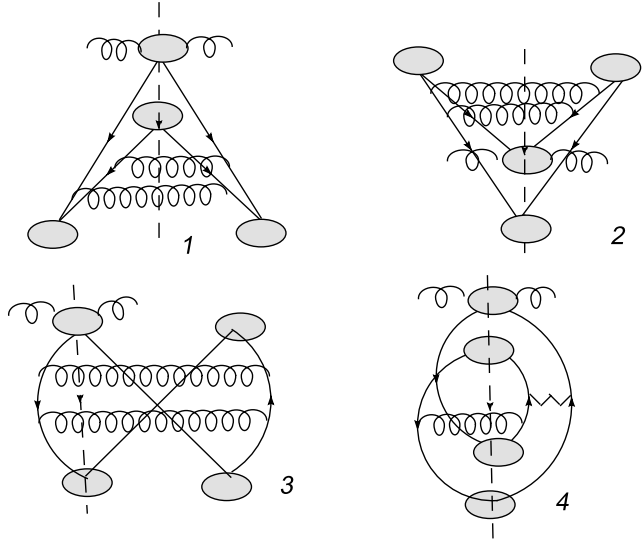}
\end{center}
\caption{Typical diagrams for emission of one gluon from the pomeron and
another from one of the two interactions with
redistribution of colour in the diffractive (1 and 2 ),
single cut (3) and double cut (4) configurations. The BKP state between the interactions is
not shown}
\label{fig7}
\end{figure}

The double inclusive cross-sections correspond to opening one of the
pomerons and one of the interactions. Some typical diagrams illustrating
contributions from various configurations are shown in Fig. \ref{fig7}
(where we suppressed $G$ between the interactions).

In the double inclusive cross-section with one of the gluons
emitted from the upper pomerons at rapidity $y_1$ we are to open either the
upper interactions, which implies fixing $Y-y'=y_2$ and the momenta
transferred to $k_2$ in one of the 4 interactions on the left in
Eq. (\ref{d3}), or the lower interactions, which implies fixing
$y''=y_2$ and the momentum transferred in one of the interactions on the
right in Eq. (\ref{d3}).

In both cases contributions will
come from DT, DP, S and DC configurations.
We shall write our results for the high-energy part $F$  with one
of the gluon emitted from the upper pomerons in the following form.
For emission of the second gluon from the interaction higher in rapidity
\beq
F^{4,hh}= \int_0^{y'}dy''
P^{(12)}_{y_1,k_1}(Y,y')P^{(34)}(Y-y')
f_{high}(y'-y'')P^{(13)}(y'')P^{(24)}(y''),\ \ y'=y_2
\label{fhh}
\eeq
and for emission of the second gluon from the interaction lower in rapidity
\beq
F^{4,hl}= \int_{y''}^{Y}dy'
P^{(12)}_{y_1,k_1}(Y,y')P^{(34)}(Y-y')
f_{low}(y'-y'')P^{(13)}(y'')P^{(24)}(y''),\ \ y''=y_2
\label{fhl}
\eeq

Operators $f_{high}$ and $f_{low}$  are different for
different configurations.
 Derivation of $f_{high}$ and $f_{low} $ can be found in ~\cite{braun2}.
One has
\beq
f_{high}=AG(y'-y'')(v_{23}-v_{34})=
\frac{1}{2}AG(y'-y'')B
\label{fhigh}
\eeq
with
the observed gluon emitted from the interaction on the left.
Remarkably $f_{low}$ is given by the same formula but with the gluon
emitted from the interaction on the right.

A similar contribution has to be added from the case
when the gluon of a higher rapidity is emitted from one of the
interactions and the one of lower rapidity from one of the lower pomerons.
This contribution is given by two
terms similar to (\ref{fhh}) and (\ref{fhl}):
for emission of the first gluon from the interaction higher in rapidity
\beq
F^{4,lh}= \int_0^{y'}dy''
P^{(12)}(Y-y')P^{(34)}(Y-y')f_{high}(y'-y'')
P^{(13)}_{y_2,k_2}(y'',0)P^{(24)}(y''),\ \ y'=y_1
\label{flh}
\eeq
and for emission of the first gluon from the interaction lower in rapidity
\beq
F_{ll}^{4,ll}= \int_{y''}^{Y}dy'
P^{(12)}(Y-y')P^{(34)}(Y-y')f_{low}(y'-y'')
P^{(13)}_{y_2,k_2}(y_1)P_{24}(y_1),\ \ y''=y_1
\label{fll}
\eeq
with the same $f_{high}$ and $f_{low}$ as given before.

Our final expression for the emission of one of the gluons
from the two explicit interactions in the diagram with the other emitted from one
of the pomerons is therefore given by the sum
\beq
F^{(4)}=F^{4,hh}+F^{4,hl}+F^{4,lh}+F^{4,ll}
\label{f4}
\eeq
with the common function
\beq
f(y'-y'')=\frac{1}{2}AG(y'-y'')B.
\label{ffin1}
\eeq
The observed gluon may be located either in the left
interaction or the right one and correspondingly $y'=y_2$ or $y''=y_2$
when the first gluon is emitted from the upper pomeron or
$y'=y_1$ or $y''=y_1$ when the second pomeron is emitted from the lower pomeron.

\subsection{Two interactions between the
pomerons  with
direct colour transmission (DCT)}
The amplitude itself is given by
(\ref{d4}) and shown im Fig. \ref{fig1},4.
In this case  the identity
of the two projectiles and two targets generates symmetries in
independent interchanges
$1\lra 2$ or $3\lra 4$.

\begin{figure}
\hspace*{2.5 cm}
\begin{center}
\includegraphics[scale=0.75]{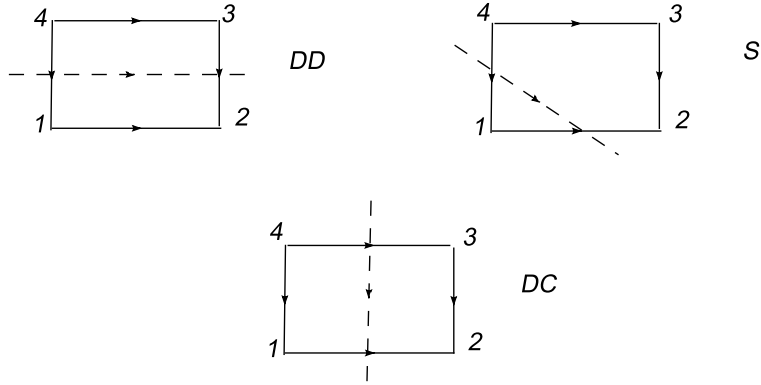}
\end{center}
\caption{
Cut interactions between the
different gluons for different cut configurations with
direct colour transmission }
\label{fig8}
\end{figure}

As in the previous subsection the relevant cuts
are to pass  through the interaction which contains the observed
gluon and also through the diagram as a whole and thus
through the BKP state. The position of the latter cut will again
be totally
determined by the cut passing through the pomerons,
and is shown  Fig. \ref{fig8} for different configurations.

\begin{figure}
\hspace*{2.5 cm}
\begin{center}
\includegraphics[scale=0.75]{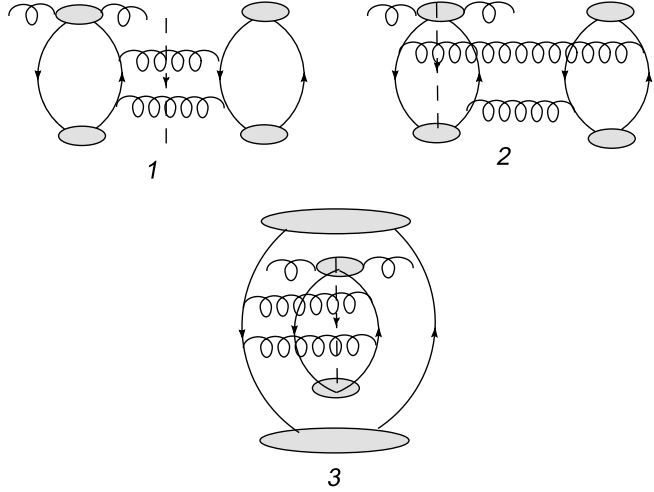}
\end{center}
\caption{Typical diagrams for emission of one gluon from the pomeron and
another from one of the two interactions with direct colour transmission
in the double diffractive (1),
single cut (2) and double cut (3) configurations. The BKP state between
the interactions is not shown}
\label{fig9}
\end{figure}

Again, since cut and uncut interactions give the same
contribution, in all cases the BKP state will appear as a whole between
the interactions which generate the observed gluon and we can
write contributions from the explicitly shown interactions
formally putting $\tilde{G}\to 1$ and
afterwards introduce this  Green
functions between the interactions.
Some typical diagrams for the double inclusive cross-sections
 corresponding to opening one of the
pomerons and one of the interactions are shown in Fig. \ref{fig9}.

As in the previous subsection the inclusive cross-section can be separated into
those in which the gluon of of higher rapidity $y_1$ is emitted from the
upper pomeron and those in which the gluon of lower rapidity $y_2$ is
emitted from the lower pomeron. In each case the other gluon may be emitted
from one of the interactions. So in the first case one has to fix in the upper(lower) interaction
$Y-y'=y_2(y'=y_2)$ and the transferred
momenta to $k_2$ and in the second case fix in the upper lower) interaction
$Y-y'=y_1 (y'=y_1)$ and the transferred momentum to $k_1$.
In both cases we write out results for the corresponding high-energy part
$F^{(5)}$  in the
same forms as before, Eqs. (\ref{fhh}), (\ref{fhl}), (\ref{flh}) and
(\ref{fll}) in which we only have to change the lower pomerons
$P^{(13)}P^{(24)}\to P^{(12)}P^{(34)}$.

The derivation of the new $f_{high}$ and $f_{low}$ can  be found
in ~\cite{braun2}.
It turns out that both are again given by the same formula
\beq
f=A\tilde{G}(y'-y'')A,
\eeq
where the gluon of higher rapidity is emitted from the left interactions
and the one of lower rapidity from the right interaction.
The total inclusive function $F^{(5)}$ will be given by a sum similar to
(\ref{f4}):
\beq
F^{(5)}=F^{5,hh}+F^{5,hl}+F^{5,lh}+F^{5,ll}
\label{f5}
\eeq

\subsection{The BKP state}
\begin{figure}
\hspace*{2.5 cm}
\begin{center}
\includegraphics[scale=0.75]{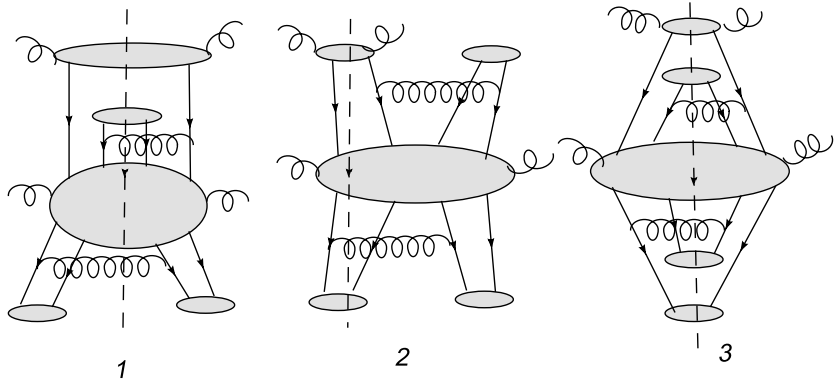}
\end{center}
\caption{Gluon production from the pomeron and the
BKP state in the diffractive (1), single (2) and double (2) cut
configurations}
\label{fig10}
\end{figure}

The simple inclusive cross-section from the BKP state was derived in ~\cite{braun2}.
So all we have to do is combine it with the inclusive cross-section from the pomeron.
With the colour redistribution the cross-section is illustrated in
Fig. \ref{fig10}. Only some typical diagtams are shown in the diffractive, single
and double cut configurations. The full list of diagrams as before can be
inferred
from ~\cite{braun2}.

We present the results in the form similar to the ones before.
For the case with colour redistribution and the gluon $(y_1,k_1)$
emitted from the upper pomeron
\beq
F^{6,h}= \int_0^Y dy'\int_0^{y'}dy''
P^{(12)}_{y_1,k_1}(Y,y')P^{(34)}(Y-y')
AG^{1243}_{y_2,k_2}(\{v_{BKP}^{1243}\},y',y'')
BP^{(13)}(y'')P^{(24)}(y'').
\label{hbkp}
\eeq
For the same case and the gluon $(y_2,k_2)$ emitted from the lower pomeron
\beq
F^{6,l}=\int_0^Y dy'\int_0^{y'}dy''
P^{(12)}(Y-y')P^{(34)}(Y-y')A
G^{1243}_{y_1,k_1}(\{v_{BKP}^{1243}\},y',y'')B
P^{(13)}_{y_2k_2}(y'',0)P^{(24)}(y'').
\label{lbkp}
\eeq
Recall that $A$ and $B$ are defined by (\ref{abdef}),
$
v_{BKP}^{1243}=v_{12}+v_{24}+v_{43}+v_{31}
$
is the sum of interactions between the neihgbouring gluons in
order 1243
and $G^{1243}_{y,k}(\{f\},y',y'')$ is given symbolically as
\beq
G^{1243}_{y,k}(\{f\},y',y'')=G^{1243}(y'-y)fG^{1243}(y-y''),
\eeq
where $f$ corresponds to emission of a gluon $(y,k)$.
The total contribution is
\beq
F^{(6)}=F^{6,h}+F^{6,l}.
\label{f6}
\eeq

For the  case without colour redistribution to obtain the
corresponding contribution $F^{(7)}$
one has to change
\[
P^{(13)}P^{(24)}\to P^{(12)}P^{(34)}\ \ {\rm and}\]\beq
AG^{1243}_{y_1,k_1}(\{v_{BKP}^{1243}\},y',y'')B\to
A[G^{1234}_{y_1,k_1}(\{v_{BKP}^{1234}\},y',y'')+
G^{1243}_{y_1,k_1}(\{v_{BKP}^{1243}\},y',y'')]A.
\label{f7}
\eeq.

\section{Both gluons emitted from the interaction}

\subsection{Two interactions with redistribution of colour}
Here we study the contribution in which one gluon is emitted from the
upper interaction in Fig. \ref{fig1},3 and the other from the lower
interaction. As before the cut BKP blob will enter as a whole and
we can formally put $G\to 1$
restoring $G$ in the final formulas.
The diagrams illustrating the double inclusive cross-section wil be the same
as in Fig. \ref{fig7} in which one should suppress emission from the
pomeron but assume instead that both interactions are opened,
that is with fixed
rapidities and transverse momenta.
The amplitude $F$ in this case can be written as
\beq
F^{(8)}= P^{(12)}(Y-y_1)
P^{(34}(Y-y_1)f^{(8)}_{k_1,k_2}P^{(13)}(y_2)P^{(24)}(y_2),
\eeq
where $f^{(8)}(k_1,k_2)$ is in fact made of products of upper and
lower interactions
in which the transferred momenta are fixed at $k_1$ for the upper interaction and $k_2$
for the lower one.
The final form of $f$ is determined after the study of D,S and DC configurations.
As compared to the previous subsection 3.1 we have to retain only
diagrams with two gluons in the intermediate state.

From the D configurations with respect to the target and
projectile we find
\[
f_{DT}=2(v_{23}+v_{14})B,\ \
f_{DP}=2A(v_{23}+v_{14}).
\]
From the S-configuration
\beq
f_{S}=
-AB-(v_{23}-v_{14})(v_{23}-v_{14}).
\label{fs2}
\eeq
The DC contributions with two gluons in the
intermediate state
is
\beq
f_{DC}= 2(v_{13}+v_{24})(v_{12}+v_{34}).
\label{DC2}
\eeq
Summing these contributions we find
\[
f^{(8)}=
AB+2(v_{23}+v_{14})(v_{23}+v_{14})-(v_{23}-v_{14})(v_{23}-v_{14}).
\]
Restoring the BKP Green function, the result can be wtitten as
\beq
f^{(8)}_{k_1,k_2}=\sum_{I,J=1}^4C_{IJ}v_I[G^{1243}(y'-y'')+
G^{1342}(y'-y'')]v_J,
\label{f8}
\eeq
where $I=(13),(24),(23),(14)$, $J=(12),(34),(23),(14)$
and coefficients $C_{IJ}$ are given in  Table 1.

\begin{center}

{\bf Table 1}

Coefficients $C_{IJ}$ with redistribution of colour

\vspace*{0.5 cm}

\begin{tabular}{|r|r|r|r|}
\hline
1&1&-1&-1 \\\hline
1&1&-1&-1\\\hline
-1&-1&2&4\\\hline
-1 &-1&4&2\\\hline
\end{tabular}
\end{center}

\subsection{Two interactions with  direct transmission  of colour}
Here we study the contribution in which one gluon is emitted from the
upper interaction in Fig. \ref{fig1},4 and the other from the lower
interaction. As before the cut BKP blob will enter as a whole and
we can study the case with $\tilde{G}\to 1$
 and introduce the full $\tilde{G}$ into the final formulas.
The diagrams illustrating the double inclusive cross-section wil be the same
as in Fig. \ref{fig9} in which one should suppress emission from the
pomeron but assume instead that both interactions are opened, that is with fixed
rapidities and transverse momenta.
The amplitude $F$ in this case can be written as
\beq
F^{(9)}= P^{(12)}(Y-y_1)
P^{(34}(Y-y_1)f^{(9)}_{k_1,k_2}P^{(12)}(y_2)P^{(34)}(y_2),
\label{f9}
\eeq
where in $f^{(9)}(k_1,k_2)$
 the transferred momenta are fixed at $k_1$ for the upper interaction and $k_2$
for the lower one.
The final form of $f^{(9)}$ is determined after the study of the
double diffractive (DD, Fig. \ref{fig9},1), S and DC
configurations.
Again we have to retain only diagrams with two gluons
in the intermediate state.

We find
\[f^{(9)}_{DD}=2(v_{13}+v_{24}-v_{23}-v_{14})(v_{13}+v_{24}-v_{23}-v_{14}),\]
\[f^{(9)}_S=
=-\frac{1}{2}\Big\{
(v_{13}-v_{14})(v_{13}-v_{14})+(v_{24}-v_{14})(v_{24}-v_{14})+
(v_{23}-v_{24})(v_{23}-v_{24})+(v_{23}-v_{13})(v_{23}-v_{13})\Big\},\]
\[f^{(9)}_{DC}
=(v_{23}+v_{14})(v_{23}+v_{14})+(v_{24}+v_{13})(v_{24}+v_{13}).\]

Summing these contributions and restoring the BKP Green functions we find
\beq
f^{(9)}_{k_1,k_2}=\frac{1}{2}\sum_{I,J=1}^4\tilde{C}_{IJ}v_I\tilde{G}(y'-y'')v_J,
\label{ff9}
\eeq
where
 $I,J=(13),(24),(23),(14)$
and coefficients $\tilde{C}_{IJ}$ are given in  Table 2.

\begin{center}

{\bf Table 2}

Coefficients $\tilde{C}_{IJ}$ with direct colour transmission

\vspace*{0.5 cm}

\begin{tabular}{|r|r|r|r|}
\hline
4&6&-3&-3 \\\hline
6&4&-3&-3\\\hline
-3&-3&4&6\\\hline
-3&-3&6&4\\\hline
\end{tabular}
\end{center}

\subsection{One gluon from the interaction, the other from the BKP state.
Redistribution of colour}
The amplitude $F$ in this case can be written as
\beq
F^{(10)}=P^{(12)}(Y-y_1)
P^{(34}(Y-y_1)f^{(10)}_{k_1,k_2}P^{(13)}(y_2)P^{(24)}(y_2).
\label{f10}
\eeq
It can be illustrated by the same diagrams as in Fig \ref{fig10} with
the emission of the pomeron suppressed but instead one of the two interactions
opened.
We have two possibilities: either the guon of higher rapidity is emitted by the
upper interaction and the one of lower rapidity from the BKP state and
{\it vice versa}.

We start from the first case.
All we have to do is to take the known inclusive cross-sections from the upper interaction
and combine them with also known inclusive cross-sections from the BKP state.
However we have to do it separately for each configuration of the cutting plane
and also take into account the symmetry factors and the minus factor for the
single cut for the contribution for a given configuration as a whole.

In the DT configuration from the interaction and from the BKP state we have
respectively
$V_{DT}=2(v_{23}+v_{14})B$ and
$
BKP_{DT}=v_{13}+v_{24}$.
This gives a contribution to $f$
$f_{DT}=2(v_{13}+v_{24})G(v_{13}+v_{24})GB$
where $G$ is either $G^{1243}$ or $G^{1342}$ with weight 1/2.
Suppressing the common riight factor $GB$ and
 the remainig Green function we rewrite it as
$f_{DT}=2(v_{13}+v_{24})(v_{13}+v_{24})$.
In the DP  configuration
from the interaction and from the BKP state we have
respectively
$V_{DP}=AB$ and
$
BKP_{DP}=v_{12}+v_{34}.
$
This gives a contribution
$f_{DP}=2A(v_{12}+v_{34})$.
The single cut contribution from the interaction (without the minus sign) is
$V_{S}=-(v_{13}-v_{14})B$
and from the BKP state
$BKP_S= v_{12}+v_{13}$.
This gives a contribution for a given choice of participants
$S=(v_{13}-v_{14})(v_{12}+v_{13})$.
The total contribution from the S configuration will be given by
\[
f_S=
(v_{13}-v_{14})(v_{12}+v_{13})+(v_{13}-v_{23})(v_{13}+v_{34})\]\[
+(v_{24}-v_{23})(v_{12}+v_{24})+(v_{24}-v_{14})(v_{24}+v_{34}).
\]
Finally the DC contribution from the interaction and the BKP state is
respectively
$V_{DC}=-(v_{13}+v_{24})B$ and
$BKP_{DC}=v_{BKP}^{1243}$,
so the total contributions is
$f_{DC}=-(v_{13}+v_{24})v_{BKP}^{1243}$.

In the sum the total double inclusive $f^{10,h}$ can be written as
\beq
f^{10,h}_{k_1,k_2}=\frac{1}{2}\sum_a\sum_{I,J=1}^4D^h_{IJ}v_IG^{a}v_JG^{a}B,
\eeq
where $a=1243,1342$, $I=(23),(14),(13),(24)$, $J=(12),(24),(34),(13)$
and the coefficients $D^h$ are given in  Table 3.

\begin{center}

{\bf Table 3}

Coefficients $D^h_{IJ}$ with redistribution of colour

\vspace*{0.5 cm}

\begin{tabular}{|r|r|r|r|}
\hline
1&1&1&1 \\\hline
1&1&1&1\\\hline
-3&-2&-3&0\\\hline
-3&0&-3&-2\\\hline
\end{tabular}
\end{center}

The case of the gluon of the lower rapidity emitted from the interaction
is considered quite similarly. The corresponding double inclusive
function $f^{10,l}$ can be written as
\beq
f^{10,l}_{k_1,k_2}=\frac{1}{2}\sum_a\sum_{I,J=1}^4D^l_{IJ}AG^{a}v_IG^{a}v_J
\eeq
where $a=1243,1342$, $I=(12),(24),(34),(13)$, $J=(23),(14),(12),(34)$
and the coefficients $D_h$ are given in  Table 4.

\begin{center}

{\bf Table 4}

Coefficients $D^l_{IJ}$ with redistribution of colour

\vspace*{0.5 cm}

\begin{tabular}{|r|r|r|r|}
\hline
1&1&0&-2 \\\hline
1&1&-3&-3\\\hline
1&1&-2&0\\\hline
1&1&-3&-3\\\hline
\end{tabular}
\end{center}
 In the end the total inclusive function $F^{(10)}$ will be given by
(\ref{f10}) with $f^{(10)}=f^{10,h}+f^{10,l}$.

\subsection{One gluon from the interaction, the other from the BKP state.
Direct transmission of colour}

The amplitude $F$ in this case can be written in the form
\beq
F^{(11)}= P^{(12)}(Y-y_1)
P^{(34}(y-y_1)f^{(11)}_{k_1,k_2}P^{(12)}(y_2)P^{(34)}(y_2)
\label{f11}
\eeq
The relevant BKP Green function is one of $G^{1234},G^{1432},G^{1243}$ and
$G^{1342}$ with weight 1/4.
Again we have two possibilities: either the guon of higher rapidity is emitted by the
upper interaction and the one of lower rapidity from the BKP state and {\it vice versa}.
We start from the first case.

We find for the DD contribution
from the interaction
$V_{DD}=AA$
and from the BKP state
$BKP=v_{14}+v_{23}$.
Using our previous notations and suppressing the right $GA$ we have
the contribution
$f_{DD}=2A(v_{14}+v_{23})$
For the S contribution we find from the interaction
$V_{S}=-(v_{13}-v_{14})$
(without the minus sign). From the BKP state we find
$BKP=v_{12}+v_{14}$, which gives
the contribution to $f$
\[
f_S=(v_{13}-v_{14})(v_{12}+v_{14})+(v_{24}-v_{14})(v_{34}+v_{14}).
\]
Finally for the DC contribution we find from the interaction
$ V_{DC}= v_{23}+v_{14}$
and from the BKP state
$BKP=v_{12}+v_{34}$,
which gives the contribution to $f$
$f_{DC}=2(v_{23}+v_{14})(v_{12}+v_{34})$.

Summing all the contributions we find
\beq
f^{11,h}_{k_1,k_2}=\frac{1}{4}\sum_a
\sum_{I,J=1}^4\tilde{D}^h_{IJ}v_IG^{a}v_JG^{a}A
\eeq
where $a=1234,1432,1243,1342$, $I=(23),(14),(13),(24)$, $J=(12),(23),(34),(14)$
and the coefficients $\tilde{D}_h$ are given in  Table 5.

\begin{center}

{\bf Table 5}

Coefficients $\tilde{D}^h_{IJ}$ with direct transmission of colour

\vspace*{0.5 cm}

\begin{tabular}{|r|r|r|r|}
\hline
2&2&2&2 \\\hline
1&2&1&0\\\hline
1&-2&0&-1\\\hline
0&-2&1&-1\\\hline
\end{tabular}
\end{center}

In the case when the gluon of the lower rapidity is emitted from the
interaction we similarly find
\[f_{DD}=2(v_{14}+v_{23})A,\]
\[f_S=(v_{12}+v_{14})(v_{13}-v_{14})+(v_{34}+v_{14})(v_{24}-v_{14}),\]
\[f_{DC}=2(v_{12}+v_{34})(v_{23}+v_{14}).\]
Summed this gives
\beq
f^{11,l}_{k_1,k_2}=
\frac{1}{4}\sum_a\sum_{I,J=1}^4\tilde{D}^l_{IJ}AGv_IGv_J,
\eeq
where $a=1234,1432,1243,1342$, $I=(12),(23),(34),(24)$, $J=(23),(14),(13),(24)$,
and the coefficients $\tilde{D}_l$ are given in Table 6.

\begin{center}

{\bf Table 6}

Coefficients $\tilde{D}^l_{IJ}$ with direct transmission of colour

\vspace*{0.5 cm}

\begin{tabular}{|r|r|r|r|}
\hline
2&1&1&0 \\\hline
2&2&-2&-2\\\hline
2&1&0&1\\\hline
2&0&-1&-1\\\hline
\end{tabular}
\end{center}
The final inclusive function $F^{(11)}$ will be given by
(\ref{f11}) with $f^{(11)}=f^{11,h}+f^{11,l}$.

\subsection{Both interactions from the BKP state.}
\begin{figure}
\hspace*{2.5 cm}
\begin{center}
\includegraphics[scale=0.75]{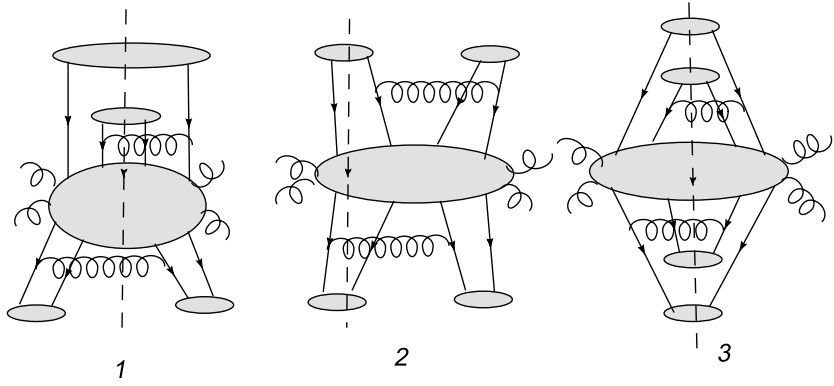}
\end{center}
\caption{Double gluon production the
BKP state in the diffractive (1), single (2) and double (2) cut
configurations}
\label{fig11}
\end{figure}

When both gluons are emitted from the BKP state, similarly to the
BFKL pomeron, we have to 'open'
it twice, that is change
\beq
G\to Gf_{y_1,k_1}Gf_{y_2,k_2}G.
\eeq
It is important that the cutting plane cannot change its position inside the BKP state,
so that function $f$ has to be the same in both 'openings,
but different for different cuts.
The double inclusive cross-section is illustrated by typical diagrams shown
in Fig. \ref{fig11}.

Let us first consider the case of colour redistribution.
The relevant BKP Green function is one of $G^{1243}$ and $G^{1342}$.

The amplitude $F$ in this case can be written as
\beq
F^{(12)}= P^{(12)}(Y-y_1)
P^{(34}(y-y_1)Af^{(12)}_{k_1,k_2}BP^{(13)}(y_2)P^{(24)}(y_2)
\label{f12}
\eeq

In the DT configuration we find the 'opened' Green function
\[f_{DT}=2G(v_{12}+v_{34})G(v_{12}+v_{34)}G.\]
In the DP configuration
\[f_{DP}=2G(v_{13}+v_{24})G(v_{13}+v_{24)}G.\]
In the S-configuration
\[f_S=-G(v_{12}+v_{13})G(v_{12}+v_{13})G-G(v_{13}+v_{34})G(v_{13}+v_{34})G\]\[
-G(v_{12}+v_{24})G(v_{12}+v_{24})G-G(v_{34}+v_{24})G(v_{34}+v_{24})G.\]
In the DC configuration
\[ f_{DC}=2Gv_{BKP}^{1243}v_{BKP}^{1243}G.\]

The final $f$ can be presented as
\beq
f^{(12)}_{k_1,k_2}=
\frac{1}{2}\sum_a\sum_{I,J=1}^4E_{IJ}G^{a}v_IG^{a}v_JG^{a},
\eeq
where $a=1243, 1342$, $I,J=(12),(24),(34),(13)$
and the coefficients $E$ are given in  Table 7.

\begin{center}

{\bf Table 7}

Coefficients $E_{IJ}$ with redistribution of colour

\vspace*{0.5 cm}

\begin{tabular}{|r|r|r|r|}
\hline
2&1&4&1\\\hline
2&2&1&4\\\hline
4&1&2&2\\\hline
1&4&1&2\\\hline
\end{tabular}
\end{center}

Now the case of direct colour transmission. The BKP Green funcion $G$
is one of four $G^{1234},G^{1432},G^{1243}$ and
$G^{1342}$ with weight 1/4.

The inclusive function ${F}$ in this case can be written as
\beq
F^{(13)}=\int d\tau P^{(12)}(Y-y_1)
P^{(34}(y-y_1)Af^{(13)}_{k_1,k_2}AP^{(12)}(y_2)P^{(34)}(y_2).
\label{f13}
\eeq

In the DD configuration we find the 'opened' Green function
\[f_{DD}=2G(v_{14}+v_{23})G(v_{14}+v_{23)}G.\]

In the S-configuration
\[f_S=-G(v_{12}+v_{14})G(v_{12}+v_{14})G
-G(v_{14}+v_{34})G(v_{14}+v_{34})G.
\]

In the DC configuration
\[ f_{DC}=2G(v_{12}+v_{34})v_{12}+v_{34})G.\]

The final $f$ can be presented as

\beq
f^{(13)}_{k_1,k_2}=\frac{1}{4}\sum_a
\sum_{I,J=1}^4\tilde{E}_{IJ}G^{a}v_IG^{a}v_JG^{a}
\eeq
where $a=1234,1432,1243,1342$, $I,J=(12),(23),(34),(14)$
and the coefficients $\tilde{E}$ are given in  Table 8.

\begin{center}

{\bf Table 8}

Coefficients $\tilde{E}_{IJ}$ with direct transmission of colour

\vspace*{0.5 cm}

\begin{tabular}{|r|r|r|r|}
\hline
1&0&2&-1\\\hline
0&2&0&2\\\hline
2&0&1&-1\\\hline
-1&2&-1&0\\\hline
\end{tabular}
\end{center}

\section{Conclusions}
We have calculated the non-trivial part of the double inclusive
cross-section to produce two gluon jets in collision of two projectiles
on two targets. It consists of 13 terms
\beq
F=\sum_{i=1}^{13}F^{(i)},
\label{ftot}
\eeq
where particular terms are given by (\ref{f1}),(\ref{f2}), (\ref{f3}),
(\ref{f4}), (\ref{f5}), (\ref{f6}), (\ref{f7}), (\ref{f8}), (\ref{f9}),
(\ref{f10}), (\ref{f11}),
(\ref{f12}), and (\ref{f13}).
This expression is not to be considered too complicated in view of variety
of different possibilities for gluon emission. By its structure our theory
is very similar to the old Regge-Gribov model with a local pomeron and
three-pomeron interaction generalized to include four-pomeron interactions
(diagrams in Fig. \ref{fig1},1,2) and a new pomeron-like object, the BKP
state made of 4 reggeized gluons, with transitions to it from normal
pomerons. Obviously the two gluons
can be emitted from old and new pomerons and also from the two
transition vertices. This gives rise to a multitude of contributions, which
lead to (\ref{ftot}).

The found inclusive cross-section depends on three rapidities, the overall one
$Y$ and two rapidities $y_1>>y_2$ of the observed gluons.
All  contributions have the same order
$\alpha_s^2f(\alpha_sN_c Y,\alpha_sN_c y)$ where
$y=y_1-y_2$ and $\alpha_sN_c Y\sim \alpha_sN_c y\sim 1$.
If one assumes the
standard behaviour of the BFKL pomeron at large $Y$ in accordance with the
BFKL equation, that is roughly as $\sim \exp (Y\Delta_{BFKL})$ where
$\Delta_{BFKL}=4\ln 2\alpha_sN_c/\pi$ then all our contributions obviously
grow as the pomeron squared, namely, as $\sim \exp (2Y\Delta_{BFKL})$.
In this limit the BKP state appears at finite rapidities, since it grows
much slowler, as $\sim \exp (Y\Delta_{BKP})$ where
$\Delta_{BKP}=0.243\Delta_{BFKL}$ ~\cite{KKM}. As to the dependence on
$y=y_1-y_2$ then if one or both  gluons are emitted from pomerons
then the contribution will grow as $\sim \exp y\Delta_{BFKL}$.
Otherwise the growth will be
much weaker, as $\sim \exp y\Delta_{BKP}$. Obviously the first type
of contributions will dominate.

However these estimates may be changed if one introduces damping of the
pomeron growth at high energies either by using the experimental
behaviour of the hadronic cross-sections or the unitarization procedure
following the Balitski-Kovchegov equation ~\cite{bal,kov}. Then the dominant
contributions will come from the BKP state with a much weaker growth
and maybe be available for experimental observation
unless the latter contributions will also be damped by absorptive
corrections similarly to the pomeron case.

Our contribution is of interest in view of
recent experimental results on long-range rapidity and azimuthal correlations
measured in experiments on colliders in nucleus-nucleus, proton-nucleus and
proton-proton collisions. The immediate
application of our results may be to deuteron-deuteron collisions.

An old question
is whether the strong azimuthal decorrelation found in the strict application
of the BFKL equation to dijet production
(Mueller-Navelet jets, ~\cite{mulnav}) is softened when the equation is
generalized to include new contributions or higher order corrections.
In particular in ~\cite{colfe} it was found that next-to-leading order
corrections to the BFKL equation, which are known to be quite large,
drastically diminish decorrelation, so that the azimuthal asymmetry in
dijet production remains well preserved up to rapidity distance $y= 10$.
In our paper we have actually studied a different sort of corrections,
of the leading order in $\alpha_sN_c$ but subdominant in $1/N_c^2$, which
involve quite different diagrams. Staying at leading order in $\alpha_sN_c$
we find that all parts of the double inclusive cross-section which involve
emission from the pomeron will damp azimuthal asymmetry exactly as
for the simple BFKL chain, that is roughly as $\exp(-y\Delta_{BFKL})$,
The new element is, however, inclusion of the intermediate BKP state
whose ground state has a much smaller intercept $\Delta_{BKP}$.
Since this ground state is non-degenerate it is azimuthally symmetric.
If, as with the BFKL pomeron, the lowest intercept of the azimuthally
anisotropic states lies around zero then azimuthal
decorrelation of jet pairs emitted
from the BKP state will be much weaker than from the BFKL pomeron.
In fact with $\Delta_{BFKL}=0.3$ the decorrelation factor
$\exp (-y\Delta_{BKP})$ is around 0.5 even at rapidity distance as high as
$y=10$. This effect can be traced experimentally even if the relative
contribution from the BKP state is small.
Naturally this conclusion has to be checked by the study of the
spectrum of the BKP states with non-zero angular momentum, which is unknown
at present. This problem is postponed for future investigation.
As mentioned in the Introduction the origin of azimuthal asymmetry
in nucleus-mucleus collisions due to diagrams like Fig. \ref{fig1},3,4
was considered in ~\cite{double}. However the possibility of the
intermediate BKP state between the interactions and the following
(possibly weak) azimuthal decorrelation was not taken into account there.

\section{Acknowledgments}

This work has been supported by the RFFI grant 12-02-00356-a
and the SPbSU grants 11.059.2010, 11.38.31.2011 and 11.38.660.2013.
The author is indebted to J.Kotanski for discussions about the properties
of the BKP states.



\begin{thebibliography}{100}
%
\bibitem{braun1} M.A.Braun, Eur. Phys. J. {\bf C 73} (2013) 2418;
arxiv:1301.4846 [hep-ph]
%
\bibitem{braun2} M.A.Braun, Eur. Phys. J. {\bf C 73} (2013) 2511;
arxiv:1305.1712 [hep-ph]
%
\bibitem{bartels} J.Bartels, Nucl. Phys. {\bf B 175} (1980) 365.
%
\bibitem{kwie} J.Kwiecinski, M.Praszalowicz,
Phys. Lett. {\bf B 94} (1980) 413.
%
\bibitem{dusling} K.Dusling, F.Gelis, T. Lappi, R.Venugopalan, Nucl. Phys.
{\bf A 836} (2010) 159.
%
\bibitem{double} A.Dumitru, J. Jalilian-Marian, Phys. Rev.
 {\bf D 81} (2010) 094015.
%
\bibitem{KKM} G.P.Korchemsky, J.Kotansky and A.N. Manashov, Phys. Rev. Lett.
{\bf 88} (2002) 122002.
%
\bibitem{bal} I.Balitski, Nucl. Phys. {\bf b 463} (1996) 99.
%
\bibitem{kov} Yu. Kovchegov, Phys. Rev. {\bf D 60} (1999) 034008.
%
\bibitem{mulnav} A.H.Mueller and H.Navelet, Nucl. Phys.
{\bf B 282} (1987) 727.
%
\bibitem{colfe} D.Colferai,F.Schwennsen, L.Szymanovski and S.Wallon,
JHEP {\bf 1012},(2010) 026.

\end{thebibliography}
\end{document}